\newcommand{\titulo}{Evolution of the Cosmological Horizons in a Concordance Universe}
\newcommand{\tema}{Cosmology}

\documentclass[10pt,a4paper]{article}

\usepackage{amsmath}                    % Matemáticas, poner separado del resto (incompatibilidad de no sé qué)
\usepackage{amsfonts,amssymb,amsthm}    % Para escribir en Matemáticas
\usepackage[english]{babel}                      % Castellaniza todo (nombres de entornos, podemos poner acentos y la ñ...)
\usepackage[latin1]{inputenc}
\usepackage[breaklinks=true]{hyperref}  % Hay que añadir las opciones luego
\usepackage[T1]{fontenc}
\usepackage{appendix}
\usepackage[nottoc]{tocbibind}          % Añade la bibliografía, lista de tablas, imágenes..

\usepackage{geometry}
\usepackage[width=.85\textwidth,font=small,labelfont=bf,labelsep=endash]{caption}

\usepackage{graphicx}
      \graphicspath{{Imagenes}}

\usepackage{smartref}
      \addtoreflist{section}   % No siempre es necesario
      \newcommand{\refconchap}[1]{\hyperref[#1]{\textcolor{\corlink}{\sectionref{#1}.}\ref*{#1}}}

\usepackage{fancyhdr}
\allowdisplaybreaks
    \newcommand{\corurl}{magenta}
    \newcommand{\corcite}{red}
    \newcommand{\corlink}{blue}
    \newcommand{\corfile}{black}

    \hypersetup{linktocpage,colorlinks,urlcolor=\corurl,citecolor=\corcite,linkcolor=\corlink,filecolor=\corfile, pdfnewwindow,pdftitle={\titulo},pdfauthor={Berta Margalef-Bentabol Juan Margalef-Bentabol},pdfsubject={\tema}}

    \theoremstyle{definition}
    \newtheorem*{remarks}{Remarks}
    \newtheorem*{properties}{Properties}
    \newtheorem*{lemmas}{Lemmas}
    \newtheorem{theorem}{Theorem}[section]

    \newtheorem{definition}[theorem]{Definition}
    \numberwithin{equation}{section}

    \renewenvironment{proof}[1][Proof]{ \textbf{#1.} }{\hfill\boldmath$\rule{0.5em}{0.5em}$\\}

\title{\titulo}
\author{Berta Margalef--Bentabol, Juan Margalef--Bentabol and Jordi Cepa}
\date{}

\begin{document}
    \pagestyle{fancy}
    \vspace*{2ex}

    \centerline{\textsf{\textbf{\Huge{Evolution of the Cosmological Horizons}}}}

    \mbox{}\vspace*{0.5ex}

    \centerline{\textsf{\textbf{\Huge{in a Concordance Universe}}}}

    \mbox{}\vspace*{2.5ex}

\centerline{
\begin{tabular}{ccccc}
    \textsf{\textbf{Berta Margalef--Bentabol}}{}\,$^1$ & \quad & \textsf{\textbf{Juan Margalef--Bentabol}}\,${}^{2,3}$ & \quad & \textsf{\textbf{Jordi Cepa}}\,${}^{1,4}$\\[1ex]
    \href{mailto:bmb@cca.iac.es}{bmb@cca.iac.es} & \quad & \href{mailto:juanmargalef@estumail.ucm.es}{juanmargalef@estumail.ucm.es} & \quad & \href{mailto:jcn@iac.es}{jcn@iac.es}
\end{tabular}}

\vspace*{2ex}

\begin{align*}
   {}^1 &\text{Departamento de Astrof\'isica, Universidad de la Laguna, E-38205 La Laguna, Tenerife, Spain}.\\[-0.6ex]
   {}^2 &\text{Facultad de Ciencias Matem\'aticas, Universidad Complutense de Madrid, E-28040 Madrid, Spain.}\\[-0.6ex]
   {}^3 &\text{Facultad de Ciencias F\'isicas, Universidad Complutense de Madrid, E-28040 Madrid, Spain.}\\[-0.6ex]
   {}^4 &\text{Instituto de Astrof\'isica de Canarias, E-38205 La Laguna, Tenerife, Spain.}\\
\end{align*}

\abstract{\noindent The particle and event horizons are widely known and studied concepts, but the study of their properties, in particular their evolution, have only been done so far considering a single state equation in a decelerating universe. This paper is the first of two where we study this problem from a general point of view. Specifically, this paper is devoted to the study of the evolution of these cosmological horizons in an accelerated universe with two state equations, cosmological constant and dust. We have obtained closed-form expressions for the horizons, which have allowed us to compute their velocities in terms of their respective recession velocities that generalize the previous results for one state equation only. With the equations of state considered, it is proved that both velocities remain always positive.}
\mbox{}\vspace*{2ex}

\noindent\textbf{Keywords}: Physics of the early universe -- Dark energy theory -- Cosmological simulations

\vspace*{3ex}

\begin{center}
  \begin{minipage}{95ex}
    \noindent This is an author-created, un-copyedited version of an article accepted for publication in Journal of Cosmology and Astroparticle Physics. IOP Publishing Ltd/SISSA Medialab srl is not responsible for any errors or omissions in this version of the manuscript or any version derived from it. The definitive publisher authenticated version is available online at:\vspace*{1ex}

    \hypersetup{urlcolor=blue}
    \centerline{\url{http://iopscience.iop.org/1475-7516/2012/12/035}}
    \hypersetup{urlcolor=\corurl}
    \mbox{}
  \end{minipage}
\end{center}

\setlength{\parindent}{0pt}      % Sangría al inicio de párrafo

\section{Introduction}
  In a Robertson-Walker universe, for any observer $A$ we can define two regions in the instantaneous three-dimensional space $t=t_0$. The first one is the region defined by the comoving points that have already been observed by $A$ (those comoving objects emitted some light in the past and it has already reached us), and the second one is its complement in the three-dimensional space, i.e. the region that cannot be observed by $A$ at a time $t_0$. The boundary between these two regions is the \textbf{particle horizon} at $t_0$, that defines the observable universe for $A$. Notice that the particle horizon takes into account only the past events with respect to $A$. Another horizon could be defined taking into account also the $A's$ future. This horizon is the \textbf{event horizon} and it is defined as the hyper-surface in space-time which divides all events into two classes, those that will be observable by $A$, and those that are forever outside $A's$ range of observation. This horizon determines a limit in the future observable universe \cite{Rindler}.\vspace{2ex}

  A deep study of the horizons has been made before in \cite{harrison} where only one state equation was considered. However, it is currently widely accepted that our present Universe is a \textbf{concordance universe} i.e. it is flat and dominated by two state equations (cosmological constant and dust), one of them of negative pressure, that drives the universe into an accelerated expansion. This fact leads us to make a deep study of more general situations that we summarize in a couple of papers, in this one, we make a deep study taking into account these two state equations with no curvature, so that we can obtain results applicable to the currently accepted cosmology, and in the second one \cite{margalef_countable_many}, a complete general study (at least from a mathematical point of view) has been made considering countably infinitely many state equations with or without curvature.\vspace{2ex}

    For numerical values in a concordance universe, we are going to use the following cosmological parameters:\vspace{2ex}

    \centerline{
    \begin{tabular}{|lc|cl|}\hline
      \rule{0ex}{3.2ex}Hubble constant                                         & & & $H_0=70.1\ \mathrm{km}\ \mathrm{s}^{-1}\mathrm{Mpc}^{-1}$\\[1ex]\hline
      \rule{0ex}{3.5ex}Current density of matter (dark and baryonic) parameter & & & $\Omega_{m0}=0.278$\\[1.5ex]\hline
      \rule{0ex}{3.5ex}Current density of cosmological constant parameter      & & & $\Omega_{\Lambda0}=0.722$ \\[1.2ex]
      \hline
    \end{tabular}}

    \mbox{}\vspace{1ex}

    This cosmological parameters have been obtained from \cite{parametros} where a combination of data from WMAP, BAO and SNCONST is considered. Note that as a good first approximation, we are neglecting in this paper the parameter of density of radiation, $\Omega_{r0}$, although it should be taken into account when considering times near the beginning of the Universe, when radiation dominates over dust, as indeed we did in \cite{margalef_countable_many}.\vspace{2ex}

    This paper could be summarized as follows: in section one, we briefly introduce some basic cosmological concepts and establish the nomenclature to be used throughout the paper. In sections two and three, we define the formal concepts of particle and event horizons, and derive their integral expressions at any cosmological time. In section four we study the evolution of the horizons, obtaining more suitable expressions for the horizons at every time. Section five is devoted to gathering all the results obtained, discussion and conclusions. In this last section, some relevant graphics about the horizons and their derivatives in the at concordance universe are included. Throughout this paper, we use geometrical units where $c=G=1$.

\section{Cosmological Prerequisites}
  \subsection{Proper distance}
    In order to introduce the particle and event horizons, we need to define the \textbf{proper distance} $D_p$, which is the distance between two simultaneous events at a cosmological time $t_0$ measured by an inertial observer. Considering a homogenous and isotropic universe, we can write the Robertson-Walker metric, where $a(t)$ is the scale factor (albeit with units of longitude in our system) and $k$ the sign of the curvature:

    \begin{equation}\label{eq. RW}
        ds^2=-dt^2+a(t)^2\left(\frac{1}{1-k r^2}dr^2+r^2d\Omega^2\right)
    \end{equation}

    Isotropy of the Robertson-Walker universe allows us to consider any direction, i.e. $\theta,\phi$ constant and so $d\Omega^2=0$, and homogeneity allows us to consider that the observer is at $r=0$.  Therefore by the definition of the proper distance (where $dt=0$) we have for a given time $t$:

    \begin{equation}\label{eq. dist prop}
        D_p(R)=\int_0^{D_p} ds = a(t) \int_0^R \frac{dr}{\sqrt{1-k r^2}}
    \end{equation}

    Notice that $R$ is the radial comoving coordinate of the measured point. On the other hand, if we consider light rays $(ds=0)$, we have, from \eqref{eq. RW} (the minus sign coming from the fact that we are considering the light coming towards us):

    \begin{equation}\label{eq. dist prop 2}
        \frac{dt}{a}=-\frac{dr}{\sqrt{1-k r^2}}\qquad  \longrightarrow  \qquad\int_{t_e}^{t_o} \frac{dt'}{a(t')} = -\int_{r_e}^0 \frac{dr}{\sqrt{1-k r^2}} = \int_0^{r_e} \frac{dr}{\sqrt{1-k r^2}}
    \end{equation}

    Where the $e$ subscript stands for emission and the $o$ subscript for observation. Notice that under our assumptions $r_o=0$. Therefore, using the last two equations, the proper distance from $r=0$ to $r=R$ at a given time $t$ can be expressed as a distance measured using light as follows:

    \begin{equation}\label{def D_p}
        D_p(t_e)=a(t)\int_{t_e}^{t}\frac{dt'}{a(t')}
    \end{equation}
        
    Notice that we have exchanged the variables $R$ and $t_e$ using the biunivocal correspondence given by \eqref{eq. dist prop 2}.\vspace*{2ex}
    
\begin{figure}[ht!]\centering
     \includegraphics[width=0.5\linewidth,trim=0cm 0cm 2.15cm 1.15cm,clip]{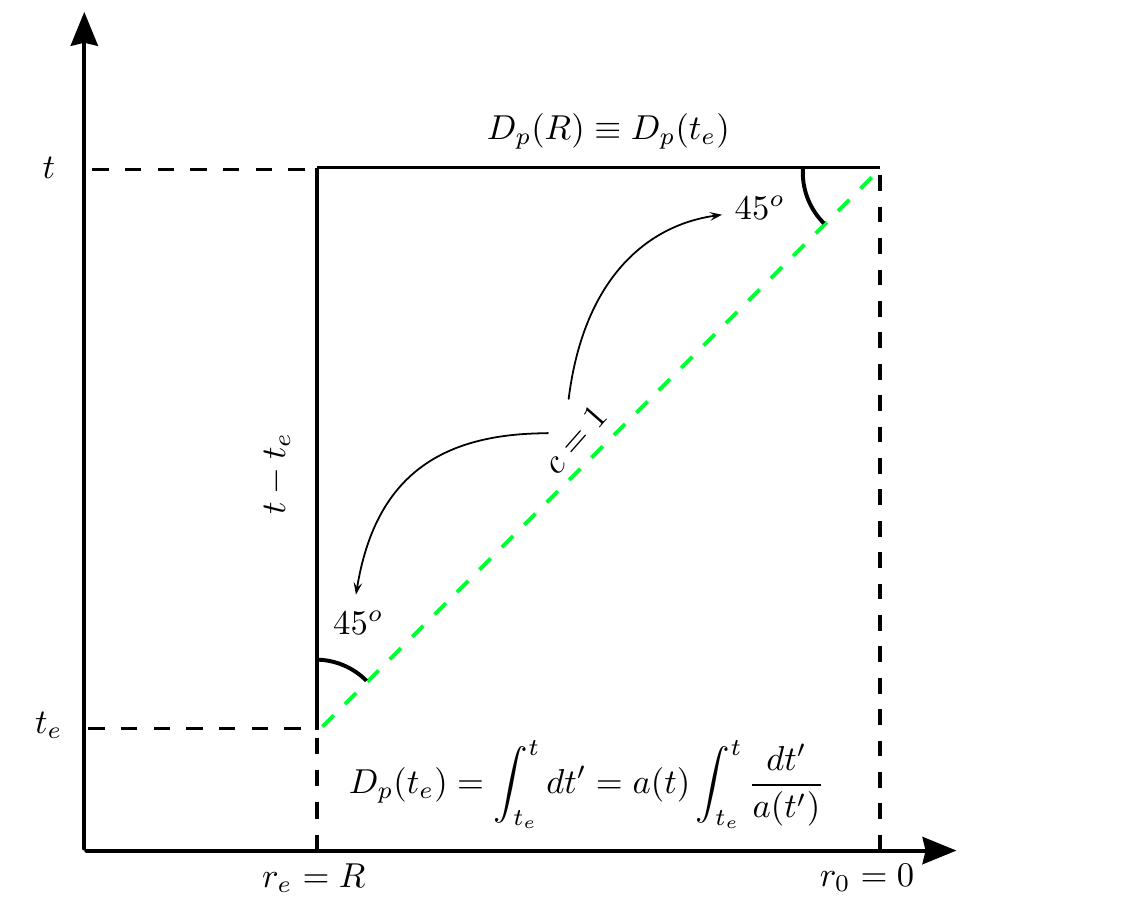}
     \caption{Representation of the measurement of the proper distance with light when $a(t)\equiv a_0$.}\label{figure_proper_time}
    \end{figure}

    So to speak, the previous formula represents the distance covered by the light between two points of the space-time, but considering an expanding universe since the $a$ factor accounts for this expansion. In particular, if $a$ is constant, then we are just measuring one cathetus using the another one and the fact that the speed of light is the same for all inertial frames (hence the angle formed is always the same) as we can see in figure \ref{figure_proper_time}.
    
  \subsection{Hubble parameter}
    The \textbf{Hubble parameter} $H$ is defined as:

    \begin{equation}\label{def H(z)}
        H(t)=\frac{1}{a}\frac{da}{dt}
    \end{equation}

    whose value at $t=t_0$ is the Hubble constant $H_0$. The product of $H$ times the proper distance, has dimensions of velocity and is known as the \textbf{recession velocity} (of a point located at comoving coordinate $R$ at time $t$):

    \begin{equation}\label{def recession}
        v_r\equiv H D_p(R)
    \end{equation}

    which physically is the instantaneous velocity of an object at a proper distance $D_p$ with respect to an inertial observer.\vspace*{2ex}

    We are considering the universe as a perfect fluid with density $\rho$ and pressure $p$, but it can also be approximated as composed by separate constituents of density $\rho_i$ and partial pressure $p_i$, where all together add up to $\rho$ and $p$ respectively. Furthermore, the $i$-th quantities are related by the linear state equation $p_i=w_i\rho_i$. Now using second Friedmann equation \cite[chap.3]{kolb1990early} (where the dot stands for time derivation): 
    
    \begin{equation}
        \dot{\rho}=-3H(\rho+p)=-3(1+w)\rho\frac{\dot{a}}{a}
    \end{equation}
    
    and considering the $i$-th state equation only, we can obtain the $i$-th density $\rho_i$ in terms of the redshift $z$:

    \begin{equation}\label{eq rho_i}
        \rho_i=\rho_{i0}(1+z)^{3\left(1+w_i\right)}
    \end{equation}

    Now we define the following time dependent magnitudes:

    \begin{itemize}
       \item[$\bullet$] Critical density as \ \ $\rho_c=\dfrac{3}{8\pi}H^2$
       \item[$\bullet$] Dimensionless energy density as \ \ $\Omega=\dfrac{\rho}{\rho_c}$
       \item[$\bullet$] Dimensionless $i$-th energy density as \ \ $\Omega_i=\dfrac{\rho_i}{\rho_c}$
    \end{itemize}

    All this quantities, when referred to the current time, are denoted with a zero subindex. If we now substitute in the definition of $\Omega_i$, the expressions of $\rho_c$, $\Omega_{i0}$ and $\rho_i$, we obtain:

    \begin{equation}\label{eq Omega_i}
        \Omega_i(z)=\Omega_{i0}H_0^2\frac{(1+z)^{3(1+w_i)}}{H(z)^2}
    \end{equation}

    Finally, adding all the $\Omega_i$ (where the index $i$ ranges over all possible state equations) leads to:

    \begin{equation}\label{H(z) sumatorio}
        H(z)=H_0\sqrt{\sum_{i=1}^n\frac{\Omega_{i0}}{\Omega}(1+z)^{3(1+w_i)}}
    \end{equation}

    \mbox{}\vspace*{-5ex}

\section{Particle Horizon}
  \subsection{At the present cosmic time}\label{subsec Hp_0}
    As the age of the Universe and the light velocity have finite values, there exists a particle horizon $H_p$, that represents the longest distance from which we can retrieve information from the past, so it defines the past observable universe. Then the particle horizon $H_p$ at the current moment is given by the proper distance measured by the light coming from $t=0$ (the origin of the universe, where a hot big bang is assumed, compatible with the dominant equation of state) to $t_0$:

    \begin{equation}\label{def H_p}
        H_p=\lim_{t_e\rightarrow  0}D_p(t_e)=\int_0^t\frac{dt'}{a(t')}
    \end{equation}

    For a rigorous discussion of the convergence of this integral see \cite{margalef_countable_many}, in this paper we will just work out our particular case. It is well known that the scale factor $a$ is related with the redshift $z$ by:

    \begin{equation}\label{def redshift}
        1+z=\frac{a_0}{a}
    \end{equation}

    Combining equations \eqref{def D_p}, \eqref{def H(z)}, the derivative of \eqref{def redshift}, and taking into account that $z(t_0)=0$, it follows that the proper distance can be written as

    \begin{equation}\label{D_p integral}
        D_p(z_e)=-\int_{z(t_e)}^{z(t_o)}\frac{dy}{H(y)}=\int_0^{z_e}\frac{dy}{H(y)}
    \end{equation}

    Particularizing equation \eqref{H(z) sumatorio} to a flat universe ($\Omega=1$) with non-relativistic matter ($w_m=0$) and cosmological constant ($w_\Lambda=-1$) \cite[chap.3]{kolb1990early}, we have:

    \begin{equation}\label{H(z)}
        H(z)=H_0\sqrt{\Omega_{m0}(1+z)^3+\Omega_{\Lambda0}}
    \end{equation}

    Applying \eqref{def H_p}, \eqref{D_p integral}, \eqref{H(z)}, and using the boundary condition that $z \rightarrow \infty$ when $t\rightarrow 0$, we obtain the expression of the particle horizon at the current cosmic time:%\enlargethispage{4ex}

    \begin{equation}\label{H_p0}
        H_p=\frac{1}{H_0}\int_0^\infty\frac{dy}{\sqrt{\Omega_{m0}(1+y)^3+\Omega_{\Lambda0}}}
    \end{equation}

    \mbox{}\vspace*{-2.8ex}

  \subsection{At any cosmic time}\label{subsec Hp(z)}
    We have so far obtained the particle horizon for the current cosmic time (corresponding to $z=0$). If we want to know the expressions of this horizon at any cosmic time, or equivalently at any other $z'$, we have to replace in \eqref{H_p0} the constant parameters $H_0$, $\Omega_{m0}$ and $\Omega_{\Lambda0}$ (corresponding to $z=0$) with the parameters corresponding to $z'$ that we denote $H(z')$, $\Omega_{m}(z')$ and $\Omega_\Lambda(z')$. Where we recall from equation \eqref{eq Omega_i} that $\Omega_{m}(z')$ and $\Omega_\Lambda(z')$ are the dimensionless energy densities given by:

    \begin{equation}\label{Omega_i(z)}
        \Omega_\Lambda(z')=\Omega_{\Lambda0}\frac{H_0^2}{H(z')^2}\qquad \qquad \Omega_m(z')=\Omega_{m0}\frac{H_0^2}{H(z')^2}(1+z')^3
    \end{equation}

    Therefore, we obtain the particle horizon for any redshift $z$ (where the prime is omitted in order to simplify the notation):
    \begin{equation}\label{H_p(z) integral}
        H_p(z)=\frac{1}{H(z)}\int_0^\infty\frac{dy}{\sqrt{\Omega_{m}(z)(1+y)^3+\Omega_{\Lambda}(z)}}
    \end{equation}

\section{Event Horizon}
  \subsection{At the present cosmic time}
    The event horizon represents the barrier between the future events that can be observed, and those that cannot. It sets up a limit in the future observable universe, since in the future the observer will be able to obtain information only from events which happen inside their event horizon. According to its definition the event horizon can be expressed as:

    \begin{equation}\label{def H_e}
        H_e=\lim_{t\rightarrow  \infty}(-D_p)
    \end{equation}

    Proceeding as in section \ref{subsec Hp_0}, but now taking into account that $z\rightarrow-1$ as $t\rightarrow \infty$, we obtain that the event horizon is:

    \begin{equation}\label{H_e integral}
        H_e=\frac{1}{H_0}\int_{-1}^0\frac{dy}{\sqrt{\Omega_{m0}(1+y)^3+\Omega_{\Lambda0}}}
    \end{equation}

  \subsection{At any cosmic time}
    Now proceeding as in section \ref{subsec Hp(z)}, the event horizon at any $z$ is:

    \begin{equation}\label{H_e(z)}
        H_e(z)=\frac{1}{H(z)}\int_{-1}^0\frac{dy}{\sqrt{\Omega_{m}(z)(1+y)^3+\Omega_{\Lambda}(z)}}
    \end{equation}

\section{Study of the Evolution of the Horizons}
  In this section we are going to obtain the expressions of particle and event horizons (as well as its derivatives) using hypergeometric functions. We would like to recall that not only is this approach interesting because we obtain analytical expressions for the horizons, but it is also a great advantage since these functions have been widely used and analysed.

  \subsection{Mathematical background}
    The theory of hypergeometric functions can be found in many books devoted to advanced calculus. For general theory see for example \cite{andrews} or \cite[chap.~9]{lebedev}, the latter is the one we are following here. In what follows, we are going to establish the minimum amount of theory that we need for our purposes.\vspace{2.5ex}

    \begin{definition}\mbox{}\\
      $_2F_1$ has the following (integral) definition:

      \begin{equation}\label{def_2F1}
        _2F_1(\alpha,\beta;\gamma;x)=\frac{\Gamma(\gamma)}{\Gamma(\beta)\Gamma(\gamma-\beta)}\int_0^1\frac{t^{\beta-1}(1-t)^{\gamma-\beta-1}}{(1-tx)^\alpha}dt \qquad
        \begin{array}{l}
            Re(\gamma)>Re(\beta)>0 \\
            |arg(1-x)|<\pi
        \end{array}
      \end{equation}
    \end{definition}

    where $\Gamma(z)$ is the gamma function, defined by:

    \begin{equation}\label{def_gamma}
        \Gamma(z)=\int_0^\infty e^{-t}t^{z-1} dt,\qquad Re(z)>0
    \end{equation}

    and by analytic continuation for every $z\in\mathbb{C}$ apart from $z=0,-1,-2\ldots$ where simple poles appears.\vspace*{3ex}

    \begin{remarks}\mbox{}\renewcommand{\labelenumi}{\thesection.\arabic{enumi}}
      \begin{enumerate}
        \setcounter{enumi}{\value{theorem}}
        \item Let us recall the important factorial property of the Gamma function, $\Gamma(z+1)=z\Gamma(z)$ for every $z\in\mathbb{C}$ apart from $z=0,-1,-2\ldots$ and $\Gamma(1)=1$.
        \item Since no distinction between hypergeometric functions $_pF_q$ is necessary, from now on, we are going to write ${}_2F_1\equiv F$.
        \item The function $F$, as a function of $x$, is defined in the whole complex plane cut along $[1,\infty)$, though we would only need real negative values.
        \item The function $F$ can be extended for every $\alpha$, $\beta$ $\in$ $\mathbb{C}$ and for every $\gamma\neq 0,-1,-2\ldots$ through recursive equations, in fact, $F(\alpha,\beta;\gamma;x)/\Gamma(\gamma)$ is an entire function of the parameters $\alpha,\beta,\gamma$. For more details see \cite[sect.~9.4]{lebedev}.
        \item When we restrict $F$, considered as a function of $x$ with fixed real parameters, to real values of $x$, we obtain a real-valued function $F:(-\infty,1)\rightarrow\mathbb{R}$.
        \item Applying the properties of the beta function $B(x,y)$\\[2ex]
            \centerline{$\displaystyle B(x,y)=\int_0^1 t^{x-1}(1-t)^{y-1}dt,\qquad \ Re(x),Re(y)>0$}
            we can rewrite $\displaystyle\frac{\Gamma(\gamma)}{\Gamma(\beta)\Gamma(\gamma-\beta)}=\frac{1}{B(\beta,\gamma-\beta)}$.\vspace*{1ex}
        \setcounter{theorem}{\value{enumi}}
      \end{enumerate}
    \end{remarks}

    Now we are going to state some useful properties (always considering that the third argument is different from $0,-1,-2\ldots$ and the fourth one verifies $|arg(1-x)|<\pi$) that can be found in \cite{andrews} or in \cite[sect.~9.2--9.8]{lebedev}.\vspace*{3ex}

    \begin{properties}\mbox{}\mbox{}\renewcommand{\labelenumi}{\thesection.\arabic{enumi}}
      \begin{enumerate}
        \setcounter{enumi}{\value{theorem}}
        \item\ $F(\alpha,\beta;\gamma;0)=1$\label{prop F(z=0)}\\[0.5ex]
        \item\ $\dfrac{\partial F}{\partial x}(\alpha,\beta;\gamma;x)=-\dfrac{\gamma-1}{x}\biggl[F(\alpha,\beta;\gamma;x)-F(\alpha,\beta;\gamma-1,x)\biggr]$\label{prop dF/dx}\\[0.7ex]  % 9.2.13 pag 243
        \item\ $\displaystyle\lim_{x\rightarrow 1^-} F(\alpha,\beta;\gamma;x)=\dfrac{B(\beta,\gamma-\beta-\alpha)}{B(\beta,\gamma-\beta)}\qquad Re(\gamma-\beta-\alpha)>0\label{prop F(z=1)}$\\[1.2ex]
        \item\ $F(\alpha, \beta;\beta;x)=(1-x)^{-\alpha}$\label{prop (a,b,b)}\\[0.5ex] % 9.8.1 pag 258
        \item\ $F(\alpha, \beta;\gamma;x)=\dfrac{1}{(1-x)^\alpha}F\left(\alpha,\gamma-\beta;\gamma;\dfrac{x}{x-1}\right)$\label{prop. ext analitica}\\[0.9ex] % 9.5.1 pag 247
        \item\ $\displaystyle \lim_{x \rightarrow \infty}F(\alpha,\beta;\gamma;-x)=0 \qquad \begin{array}{l}
            \alpha,\beta>0\\
            \alpha-\beta\notin\mathbb{Z}
        \end{array}$\label{lim F}\\
        \setcounter{theorem}{\value{enumi}}
      \end{enumerate}
    \end{properties}

    The last expression follows by one of the analytic extensions of $F$ given in \cite[sect.~9.5]{lebedev} together with property \refconchap{prop F(z=0)}. Now we are going to obtain some particular values of $F$ that we will need.\vspace*{3ex}

   \begin{lemmas}\mbox{}\renewcommand{\labelenumi}{\thesection.\arabic{enumi}}
      \begin{enumerate}
        \setcounter{enumi}{\value{theorem}}\newcounter{hola}\setcounter{hola}{\value{theorem}}
        \item\ $\displaystyle \int_1^\infty\frac{ds}{\sqrt{s^3+A}}=2\ F\left(\frac{1}{2},\frac{1}{6};\frac{7}{6};-A\right)$ \quad if \quad $A>0$ \label{lema integral 1}\\[2ex]
        \item\ \ $\displaystyle \int_0^1\frac{ds}{\sqrt{s^3+A}}=\frac{1}{\sqrt{A}}F\left(\frac{1}{2},\frac{1}{3};\frac{4}{3};-\frac{1}{A}\right)$ \quad if \quad $A>0$\label{lema integral 2}\\[2ex]
        \item\ $\displaystyle\frac{\partial F}{\partial x}(\alpha,\beta;\beta+1;x)=-\frac{\beta}{x}\left[F(\alpha,\beta;\beta+1;x)-\frac{1}{(1-x)^{\alpha}}\right]$\label{lema derivada}\\[1ex]
        \item\ $\displaystyle \lim_{x\rightarrow 1^-}F\left(\frac{1}{2},1;1+c;x\right)=\infty$ \quad if \quad $0<c<\frac{1}{2}$\label{lema limite}
        \setcounter{theorem}{\value{enumi}}
      \end{enumerate}
    \end{lemmas}
    \mbox{}

    \begin{proof}\mbox{}\renewcommand{\labelenumi}{\thesection.\arabic{enumi}}
      \begin{enumerate}
        \setcounter{enumi}{\value{hola}}
        \item Beginning from the right hand side of the equation, substituting in the definition \eqref{def_2F1}, applying the factorial property of the Gamma function and then making the change of variable $s=t^{-1/3}$ leads to the left hand side of the equation.
        \item Analogous to the previous case, but now making the change of variable $s=t^{1/3}$.
        \item This statement follows immediately from properties \refconchap{prop dF/dx} and \refconchap{prop (a,b,b)}.
        \item Notice first that property \refconchap{prop F(z=1)} cannot be applied as the conditions on the parameters do not hold. The result follows applying property \refconchap{prop F(z=0)} to the analytical extension in $|z-1|<1$ cut along $[1,\infty)$ that can be found in \cite{lebedev}, equations (9.5.7) or (9.5.10). The hypothesis of this equations are satisfied as $c\in\left(0,\frac{1}{2}\right)$ and $x\in\mathbb{R}$.\\[-6.6ex]
      \end{enumerate}
    \end{proof}
    
    It is important to note that in the third equation, we have managed to established the derivative of $F$ (with one restriction in the parameters) in terms of the $F$ itself. Finally, we introduce the auxiliary functions $A,B$ and gather some straightforward computations that we will need to obtain the derivative of both horizons.
    \begin{align}
      \displaystyle&\qquad A\equiv \frac{\Omega_\Lambda(z)}{\Omega_m(z)}=\frac{\Omega_{\Lambda0}}{\Omega_{m0}}\frac{1}{(1+z)^3} \hspace*{7ex} \longrightarrow \ && \hspace*{-3.5ex}\frac{dA}{dz}=-\frac{3A}{1+z}\label{def A}\\[2ex]
      \displaystyle&\qquad B\equiv\frac{1}{A}=\frac{\Omega_{m0}}{\Omega_{\Lambda0}}(1+z)^3\ \hspace*{11.35ex} \longrightarrow \ && \hspace*{-3.5ex}\frac{dB}{dz}=\frac{3B}{1+z}&\\[2ex]
      \displaystyle&\qquad H(z)=H_0\sqrt{\Omega_{\Lambda0}}\sqrt{B+1}\label{H=f(B)}&\\[1.4ex]
      \displaystyle&\qquad \frac{dz}{dt}=-\frac{a_0}{a^2}\frac{da}{dt}=-\frac{a_0}{a}H=-(1+z)H(z)&\label{dz/dt}
    \end{align}

    \mbox{}\\[-0.7ex]Where the last expression is obtained applying first equation \eqref{def redshift}, then equation \eqref{def H(z)} and finally equation \eqref{def redshift} again.

  \subsection{Expressing the particle horizon through hypergeometric functions}
    The expression of the particle horizon \eqref{H_p(z) integral} can be expressed as follows, where $z$ should verify $z\in(-1,\infty)$ according to equation \eqref{def redshift}:
    \begin{align*}
      H_p(z)&=\frac{1}{H(z)}\int_0^\infty\frac{dy}{\sqrt{\Omega_m(z)(1+y)^3+\Omega_\Lambda(z)}}\overset{s=y+1}{=} \frac{1}{H(z)\sqrt{\Omega_m(z)}}\int_1^\infty\frac{ds}{\sqrt{s^3+A(z)}}\overset{\refconchap{lema integral 1}}{=}\\[2ex]
            &=\frac{2}{H(z)\sqrt{\Omega_m(z)}}F\left(\frac{1}{2},\frac{1}{6};\frac{7}{6};-A(z)\right)\overset{eq. \eqref{Omega_i(z)}}{=}\frac{2}{H_0\sqrt{\Omega_{m0}(1+z)^3}}F\left(\frac{1}{2},\frac{1}{6};\frac{7}{6};-A(z)\right)
    \end{align*}

    Finally, by definition of $A$ \eqref{def A}:

    \begin{equation}\label{H_p(hipergeom)}
        \boxed{\ \rule{0ex}{4ex} \displaystyle H_p(z)=\frac{2\sqrt{A(z)}}{H_0\sqrt{\Omega_{\Lambda0}}} F\left(\frac{1}{2},\frac{1}{6};\frac{7}{6};-A(z)\right)\ }
    \end{equation}

  \subsection{Obtaining the derivative of the particle horizon}
    Now we are going to obtain the derivative of $H_p$. Notice that in the expression of $H_p$, the parameters of $F$ verify $\gamma=\beta+1$, so we can use lemma \refconchap{lema derivada} and obtain $F$ with the same parameters. For that reason and in order to simplify the notation, we omit the parameters and we also omit the argument $z$ in the $A$ and $B$ function.
    \begin{align*}
      \frac{dH_p}{dz}&=\frac{1}{H_0\sqrt{\Omega_{m0}}}\left[2\frac{dA/dz}{2\sqrt{A}}F\left(-A\right)+2\sqrt{A} \left(-\frac{dA}{dz}\right)\frac{\partial F}{\partial x}\left(-A\right)\right]\overset{\refconchap{lema derivada}}{=}\\[2ex]
      &=\frac{1}{H_0\sqrt{\Omega_{m0}}}\left[-3\frac{\sqrt{A}}{1+z}F\left(-A\right)+    \frac{\sqrt{A}}{1+z}\left(F\left(-A\right)-\frac{1}{\sqrt{1+A}}\right)\right]\overset{B=1/A}{=}\\[2ex]
      &=-\frac{1}{(1+z)}\frac{2\sqrt{A}}{H_0\sqrt{\Omega_{m0}}}F\left(-A\right)-\frac{1}{(1+z)}\frac{1}{H_0\sqrt{\Omega_{m0}}\sqrt{1+B}}\overset{eq.\eqref{H_p(hipergeom)},\eqref{H=f(B)}}{=}\\[2ex]
      &=-\frac{H_p(z)}{1+z}-\frac{1}{(1+z)H(z)}
    \end{align*}

    Finally, applying the chain rule and equation \eqref{dz/dt}, we obtain:

    \begin{equation}\label{dH_p=f(rec)}
        \boxed{\boxed{\ \rule{0ex}{4ex} \displaystyle\frac{dH_p}{dt}=H_p(z)H(z)+1\ }}
    \end{equation}

    Note that $H_p(z)H(z)$ represents the recession velocity of the particle horizon \eqref{def recession}. Its physical meaning is explained in subsection \ref{sub conclusions}, where all the conclusions are provided.\vspace*{2ex}

  \subsection{Expressing the event horizon through hypergeometric functions}
    Analogously for $H_e$ we have:
    \begin{align*}
      H_e(z)&=\frac{1}{H(z)}\int_{-1}^0\frac{dy}{\sqrt{\Omega_m(z)(1+y)^3+\Omega_\Lambda(z)}}\overset{s=1+y}{=}\frac{1}{H(z)\sqrt{\Omega_m(z)}}\int_0^1\frac{ds}{\sqrt{s^3+A}}\overset{\refconchap{lema integral 2}}{=}\\[1.2ex]
            &=\frac{1}{H(z)\sqrt{\Omega_m(z)}}\frac{1}{\sqrt{A}}F\left(\frac{1}{2},\frac{1}{3};\frac{4}{3};-\frac{1}{A}\right)=\frac{1}{H(z)\sqrt{\Omega_\Lambda(z)}}\ F\left(\frac{1}{2},\frac{1}{3};\frac{4}{3};-\frac{1}{A}\right)
    \end{align*}

    Where in the last equality, the definition of $A$ \eqref{def A} is used. Now using equation \eqref{Omega_i(z)} we obtain the following expression:

    \begin{equation}\label{H_e(hipergeom)}
        \boxed{\ \rule{0ex}{4ex} \displaystyle H_e(z)=\frac{1}{H_0\sqrt{\Omega_{\Lambda0}}}\ F\left(\frac{1}{2},\frac{1}{3};\frac{4}{3};-B(z)\right)\ }
    \end{equation}

  \subsection{Obtaining the derivative of the event horizon}
     We are again in the hypothesis of lemma \refconchap{lema derivada}, and then it is meaningful to omit the parameters.
     \begin{align*}
        \frac{dH_e}{dz}&=\frac{1}{H_0\sqrt{\Omega_{\Lambda0}}}\left(-\frac{dB}{dz}\right)\frac{\partial F}{\partial x}\left(-B\right)\overset{\refconchap{lema derivada}}{=}\\[2ex]
        &=-\frac{1}{(1+z)}\frac{1}{H_0\sqrt{\Omega_{\Lambda0}}}F\left(-B\right)+\frac{1}{1+z}\frac{1}{H_0\sqrt{\Omega_{\Lambda0}}\sqrt{1+B}}\overset{eq.\eqref{H_e(hipergeom)},\eqref{H=f(B)}}{=}\\[2ex]
        &=-\frac{H_e(z)}{1+z}+\frac{1}{(1+z)H(z)}
     \end{align*}

     Finally, applying the chain rule and equation \eqref{dz/dt}, we obtain:

     \begin{equation}\label{dH_e=f(rec)}
        \boxed{\boxed{\ \rule{0ex}{4ex}\displaystyle\frac{dH_e}{dt}=H_e(z)H(z)-1\ }}
     \end{equation}

     Where now $H_e(z)H(z)$ is the recession velocity of the event horizon. Its physical implications will be explained in subsection \ref{sub conclusions}.

\section{Results and Interpretation}
  \subsection{Some relevant values}
    Below we show some important values of the horizons, summarized in table \ref{table relevant values}, and the required computations to obtain them.

    \begin{table}[ht!]
      \centering
      \begin{tabular}{l|c c | c c|c c c c}
        \hline
        \rule{0ex}{4.1ex} & $z$ &  $t$ & $\Omega_m(z)$ & $\Omega_\Lambda(z)$ & $H_p(z)$ & $H_e(z)$ & $\dfrac{dH_p}{dt}$ & $\dfrac{dH_e}{dt}$\\[1.8ex]
        \hline\hline
        \rule{0ex}{3ex}Origin of the universe & $\infty$ & $0$ & $1$ & $0$ & $0$ & $0$ & $3c$ & $\infty$ \\[1.5ex]
        Current time & $0$ & $t_0$ & $0.278$ & $0.722$ & $\dfrac{2.672\,c}{H_0\sqrt{\Omega_{\Lambda0}}}$ & $\dfrac{0.958\,c}{H_0\sqrt{\Omega_{\Lambda0}}}$ & $4.409\,c$ & $0.128\,c$\\[3ex]
        Future &$-1$ & $\infty$ & $0$ & $1$ & $\infty$ & $\dfrac{c}{H_0\sqrt{\Omega_{\Lambda0}}}$ & $\infty$ & $0$\\[2ex]
        \hline
      \end{tabular}
      \caption{Some important values of the horizons and their velocities in physical units. Notice that the numerical values appearing on the current time values, depend on both $\Omega_{\Lambda0}$ and $\Omega_{m0}$.}\label{table relevant values}
    \end{table}

    \mbox{}\vspace*{-2.5ex}
    \begin{itemize}
      \item[$\bullet$] $H_p(\infty)\ $ -- Notice that $A \rightarrow 0$ when $z \rightarrow \infty$ and use property \refconchap{prop F(z=0)} in \eqref{H_p(hipergeom)}.\\[-1ex]
      \item[$\bullet$] $H_e(\infty)\ $ -- Notice that $B \rightarrow \infty$ when $z \rightarrow \infty$ and use property \refconchap{lim F} in \eqref{H_e(hipergeom)}.\\[-1.2ex]
      \item[$\bullet$] $\dfrac{dH_p}{dt}(\infty)\ $-- Again, $A \rightarrow 0$ when $z \rightarrow \infty$, then apply property \refconchap{prop. ext analitica} to eq. \eqref{dH_p=f(rec)}, which leads to:
           \[\frac{dH_p}{dt}=2F\left(\frac{1}{2},1;\frac{7}{6};\frac{A}{A+1}\right)+1\longrightarrow 2F\left(\frac{1}{2},1;\frac{7}{6};0\right)+1\overset{\refconchap{prop F(z=0)}}{=}3\]
      \item[$\bullet$] $\dfrac{dH_e}{dt}(\infty)\ $-- Again, $B \rightarrow \infty$ when $z \rightarrow \infty$, then apply property \refconchap{prop. ext analitica} to eq. \eqref{dH_e=f(rec)}, which leads to:
          \[\frac{dH_e}{dt}=F\left(\frac{1}{2},1;1+\frac{1}{3};\frac{B}{B+1}\right)-1\overset{\refconchap{lema limite}}{\xrightarrow{\hspace*{10ex}}}\infty\]
      \item[$\bullet$] $H_p(-1)\, $ -- Notice that $A \rightarrow \infty$ when $z \rightarrow -1$ and apply property \refconchap{prop. ext analitica} to eq. \eqref{H_p(hipergeom)}, which leads to:
      \[H_p=\frac{2}{H_0\sqrt{\Omega_{\Lambda 0}}}\sqrt{\frac{A}{A+1}}F\left(\frac{1}{2},1;1+\frac{1}{6};\frac{A}{A+1}\right)\overset{\refconchap{lema limite}}{\xrightarrow{\hspace*{10ex}}}\infty\]
      \item[$\bullet$] $H_e(-1)\ $-- Notice that $B \rightarrow 0$ when $z \rightarrow -1$ and use property \refconchap{prop F(z=0)} in \eqref{H_e(hipergeom)}.\\[-1ex]
      \item[$\bullet$] To determine the values of the derivatives for $z=-1$, just substitute $H(-1)=H_0\sqrt{\Omega_{\Lambda0}}$ in equations \eqref{dH_p=f(rec)} and \eqref{dH_e=f(rec)}, and use the values $H_p(-1)$ and $H_e(-1)$ we have just obtained above.\\[-1ex]
      \item[$\bullet$] Finally, we obtain the values at the current cosmic time through numerical calculus using \eqref{H_p(hipergeom)}, \eqref{H_e(hipergeom)}, \eqref{dH_p=f(rec)} and \eqref{dH_e=f(rec)}.
    \end{itemize}

  \subsection{The velocities are always positive}
    It is clear from equation \eqref{dH_p=f(rec)} that the velocity of the particle horizon is always positive, but from equation \eqref{dH_e=f(rec)} we cannot conclude the same for the event horizon. Let us see that with a little more effort we are able to obtain some information about the behaviour of the horizon velocities, including the fact that they are indeed always positive. In order to do that, we compute the second derivative which is quite straightforward using all the previous computations:

    \[\frac{d^2H_p}{dt^2}=H\left(\frac{2-B}{2(1+B)}H_pH + 1\right)\qquad\qquad\frac{d^2H_e}{dt^2}=H\left(\frac{2-B}{2(1+B)}H_eH - 1\right)\]

    If we equal both equations to zero, we obtain (where the correspondent parameters have been omitted):
    \begin{align*}
       \text{For } H_p: \quad F(-A)=\frac{\sqrt{1+A}}{1-2A}\qquad\qquad  \text{For } H_e: \quad F(-B)&=\frac{2\sqrt{1+B}}{2-B}
    \end{align*}

    From the definition of $F$, it is easy to prove applying elementary inequalities rules, that when $x\in(-\infty,1)$ with positive parameters the function $F(x)$ is strictly positive and increasing. Then on one hand, we have that $F(-x)$ is positive but decreasing when $x\in(-1,\infty)$.\footnote{$A$ and $B$, that by definition are both positive when $z$ ranges over $(-1,\infty)$, would play the role of the $x$.} Computing the derivative with respect to $A$ of the right hand side of first equation, we obtain on the other hand, that it is increasing where it is defined, takes positive values in the interval $I=(-1,1/2)$ and negative for larger values. Gathering these two facts we have that if there exists a solution $A_0$ for the first equation, it must be unique and $A_0\in I$. Clearly when $A=0$, the right hand side is $1$ and from property \refconchap{prop F(z=0)} $F(0)=1$, so the unique solution is $A_0=0$. Analogously for the second equation (but now $I=(-1,2)$ that does not affect the argument) we obtain that $B_0=0$ is the unique solution for the second equation.\vspace{2ex}

    Therefore, from the definition of $A$ and $B$, we have that the acceleration of $H_p$ is zero only at the origin of the universe $z\rightarrow \infty$ and the acceleration of $H_e$ is zero only at the far future $z=-1$, and for the rest of the values of $z$, both accelerations are positive.\vspace{2ex}

    As at the origin of the universe both velocities were positive, we conclude they are always nonnegative. In fact, they can only vanish for the limits $z\rightarrow\infty$ and $z\rightarrow -1$, and according to table \ref{table relevant values}, it only happens at the far future for the event horizon speed. We have included on the appendix (page \pageref{graphic Hp}) some figures that show the behaviour of both horizons and their derivative.

  \subsection{Limit to just one state equation}\label{sect. one state equation}
    In this section we are going to prove that in fact we are generalizing the previous results obtained taking into account just one state equation such as the ones stated in \cite{harrison}. In general, it can be proved that for a unique state equation the following results are obtained for any cosmic time (e.g. using \eqref{def H_p}, \eqref{D_p integral}, \eqref{H(z) sumatorio} and \eqref{eq Omega_i} for just one state equation, and \eqref{dz/dt} to compute their derivatives):

    \[\begin{array}{lclcl}
        H_p^*(z)=\displaystyle \frac{2}{H^*(z) (1+3w)}  & \quad \quad\qquad& \displaystyle \frac{dH_p^*}{dt}=\frac{3(1+w)}{1+3w}  & \qquad \text{if } w>-1/3\\[3ex]
        H_e^*(z)=\displaystyle -\frac{2}{H^*(z) (1+3w)} & \quad \quad\qquad& \displaystyle \frac{dH_e^*}{dt}=-\frac{3(1+w)}{1+3w} & \qquad \text{if } w<-1/3
    \end{array}\]

    Where the $*$ stand for the fact that we are taking into account just one state equation, in contrast with the equations obtained in this paper taking into account two state equations.\vspace{2ex}

     \begin{itemize}
        \item If it dominates the state equation of matter, we have $w=0$, then there only exists the particle horizon whose equations, as proved in \cite{harrison}, are:
            \[H_p^*(z)=\frac{2}{H^*(z)} \hspace{8ex} \frac{dH^*_p}{dt}=3\]
            In this case $\Omega_m(z)=\Omega_{m0}=1$ and $\Omega_\Lambda(z)=\Omega_{\Lambda 0}=0$, and therefore $H^*(z)=H_0\sqrt{(1+z)^3}$. Now substituting the $A$ function in equation \eqref{H_p(hipergeom)}, we have (omitting once again the parameters):
            \begin{align*}
              &H_p(z)=\frac{2}{H_0\sqrt{\Omega_{m0}(1+z)^3}} F\left(-\frac{\Omega_{\Lambda0}}{\Omega_{m0}}\frac{1}{(1+z)^3}\right)=\frac{2}{H_0\sqrt{(1+z)^3}}F\left(0\right)\overset{\refconchap{prop F(z=0)}}{=}\frac{2}{H^*(z)}\\[2.5ex]
              &\frac{dH_p}{dt}=H(z)H_p(z)+1=H^*(z)\frac{2}{H^*(z)}+1=3
            \end{align*}

        \item If it dominates the state equation of cosmological constant (i.e. $w=-1$) then there only exists the event horizon whose equations are:
            \[H_e=\frac{1}{H^*(z)} \hspace{8ex} \frac{dH_e}{dt}=0\]
            In this case $\Omega_m(z)=\Omega_{m0}=0$ and $\Omega_\Lambda(z)=\Omega_{\Lambda 0}=1$, and therefore $H^*(z)=H_0$. Then, equation \eqref{H_e(hipergeom)} leads to:
            \begin{align*}
                &H_e(z)=\frac{1}{H_0\sqrt{\Omega_{\Lambda0}}}F\left(-\frac{\Omega_{m0}}{\Omega_{\Lambda0}}(1+z)^3\right)=\frac{1}{H_0}F\left(0\right)\overset{\refconchap{prop F(z=0)}}{=}\frac{1}{H_0}=\frac{1}{H^*(z)}\\[2.5ex]
                &\frac{dH_e}{dt}=H(z)H_e(z)-1=H^*(z)\frac{1}{H^*(z)}-1=0
            \end{align*}
     \end{itemize}

     So we conclude that indeed, our results generalize the previous ones obtained in \cite{harrison}.

  \subsection{The case of the cosmic background radiation}
    The cosmic microwave background radiation that we measure today comes from a spherical surface called the \textbf{surface of last scattering}. In fact, it is not a surface but it has some thickness. In the current Standard Model, the surface of last scattering is at redshift $z_{ls}=1\,089$ with a thickness of $\Delta z=195$ \cite{spergel2003first}. The proper distance to the mean value of this surface is:

    \[D_p=\frac{1}{H_0}\int_0^{1089}\frac{dy}{\sqrt{\Omega_{m0}(1+y)^3+\Omega_{\Lambda0}}}=14\,086.37\ Mpc\]

    The values of particle and event horizons at the current cosmic time are according to table \ref{table relevant values} (with the numerical constant we mentioned at the beginning of the paper):

    \[ H_{p0}=14\,577.72\ Mpc \qquad \qquad  H_{e0}=4\,823.66\ Mpc\]

    We notice that from the integral definition of the particle horizon \eqref{H_p0}, the distance to the surface of last scattering is always smaller than the particle horizon, which agrees with the experimental data. The light which is reaching us at the current moment from objects located over the surface of last scattering was emitted at a time corresponding to a redshift of $z_{ls}$. As time goes by, the distance to the surface will be higher, but always less than the particle horizon, so in the future we get information emitted from the surface of last scattering.\vspace{2ex}

    On the other hand the distance to the last scattering surface is greater than the event horizon, which means that the light emitted at the present moment from this surface, is never going to reach us. It does not mean that in the future we are not going to see background radiation, but that the radiation we can see at the current time and which we will see in the future is the one inside the present event horizon. This can be generalized to any object in the universe, and drives to the conclusion that in an ever accelerated universe (i.e. dominated by a negative pressure) with several state equations, where the initial expansion phases were dominated by positive pressure terms, although the particle horizon guarantees that an object can be seen, the event horizon prevents to see photons beyond a time given by the instantaneous event horizon at that time. Summarizing, the particle horizon defines the events that could be observed at a given cosmic time, while the event horizon defines the events that will be observed in the future.

  \subsection{Conclusions. Recession of the cosmological horizons}\label{sub conclusions}
    In this paper, we have first obtained analytical expressions for the particle and the event horizons considering two state equations (cosmological constant and dust as accepted in the concordance cosmology). They have allowed us to compute their first derivative, obtaining these extremely simple equations:
    \[\frac{dH_p}{dt}=H_p(z)H(z)+1\qquad \qquad \frac{dH_e}{dt}=H_e(z)H(z)-1\]
    where $H_pH$ and $H_eH$ are the recession velocities of the particle and event horizons respectively. These expressions generalize the previous stated results \cite{harrison} derived for just one state equation. Notice that all these results and values are independent of the current values of the involved constants.\vspace*{2ex}

    The equations of the velocity of the horizons have a remarkable physical meaning. As the recession velocity $H_pH$ is the instantaneous velocity of an object located at the distance of the particle horizon $H_p$, from the first equation we deduce that the instantaneous velocity of the surface of the horizon particle is faster by the speed of light $c$ (in our units $c = 1$), than the one of the objects over the particle horizon, then more and more objects are entering into the particle horizon and they will never get out. Analogously, the second equation stands that the recession speed of the event horizon is slower by $c$ than that of the objects over the event horizon, and then more and more objects are disappearing from the event horizon and they will never get into again.\vspace*{2ex}

    Finally, computing the second derivatives of both horizons we have proved that under our assumptions, their velocities remain positive throughout the history of the universe.\vspace*{2ex}

    \paragraph{Acknowledgements}\mbox{}\\
      This work was partially supported by the Spanish Ministry of Economy and Competitiveness (MINECO) under the grant AYA2011-29517-C03-01.

\hypersetup{urlcolor=blue}

\newpage

\appendix

\section{Graphical Behaviour of the Horizons}\label{graphic Hp}
  In this appendix we include the graphics of the horizons and their velocities obtained with numerical calculus. As the hypergeometric functions are really well implemented in almost all mathematical programs, obtaining these graphics is extremely easy.

  \subsection{Particle horizon}
    \begin{figure}[ht!]
      \centering
      \centerline{\includegraphics[trim=3.2cm 9cm 4cm 9cm,clip,scale=0.68]{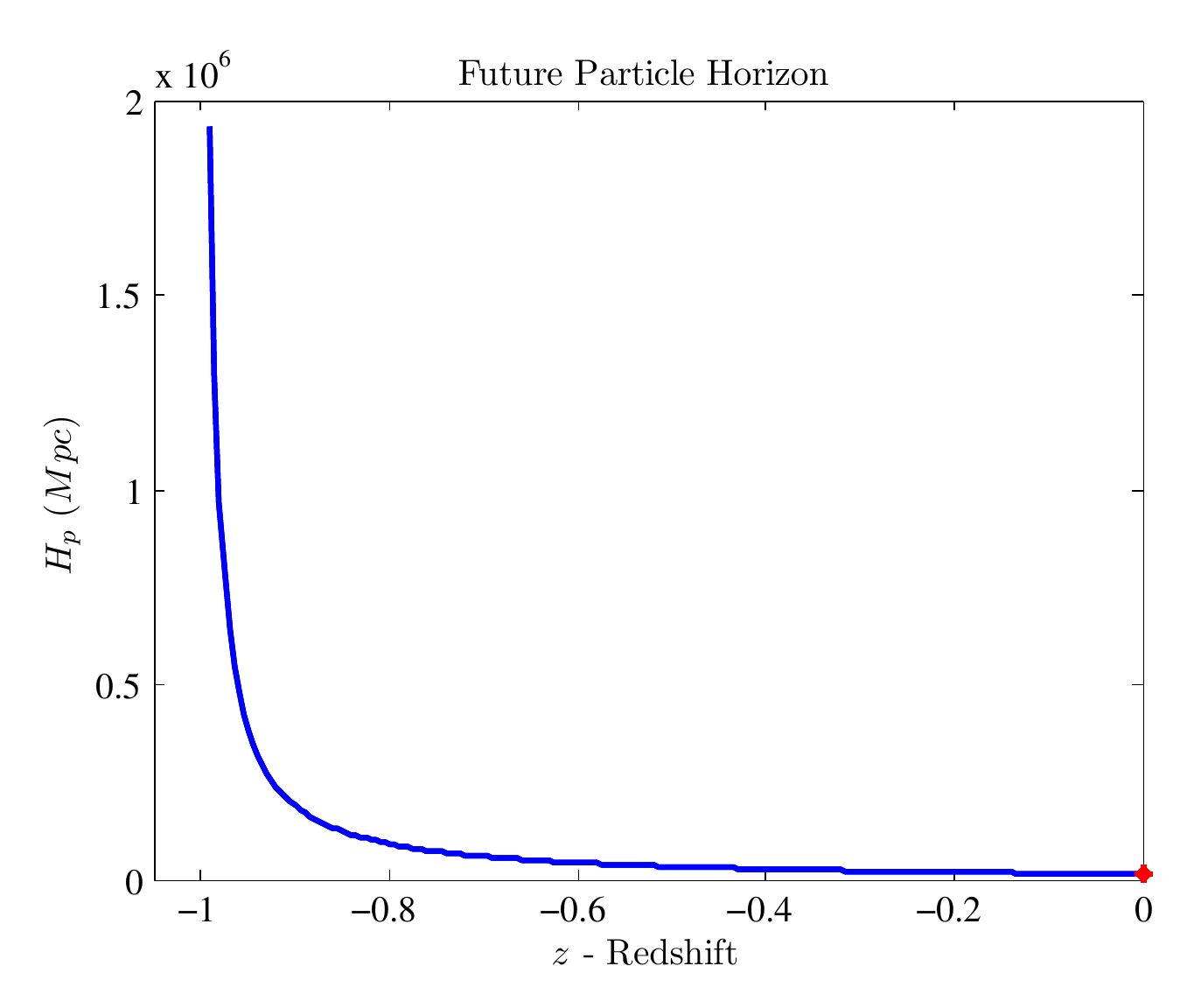} \quad \includegraphics[trim=3.2cm 9cm 4cm 9cm,clip,scale=0.68]{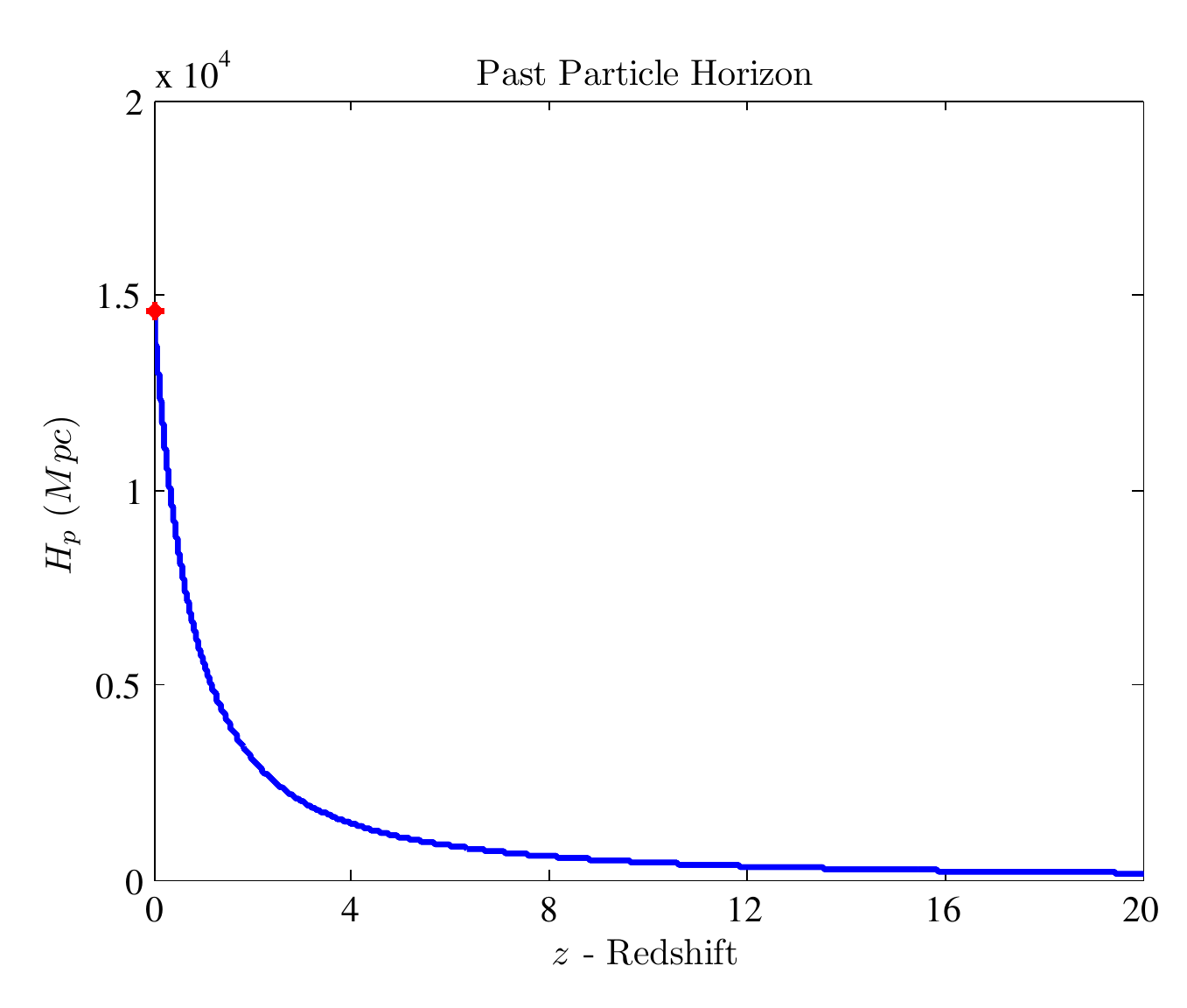}}
      \caption{Particle Horizon. At the origin of the Universe (right hand side of the graphic on the right), the particle horizon tends to zero, then it increases rapidly with time tending to infinite at the far future (left hand side of the other graphic on the left). The dot in both graphics represents the connecting point between them and corresponds to the current time $z=0$.}
    \end{figure}

    \begin{figure}[ht!]
      \centering
      \centerline{\includegraphics[trim=3.2cm 9cm 4cm 9cm,clip,scale=0.68]{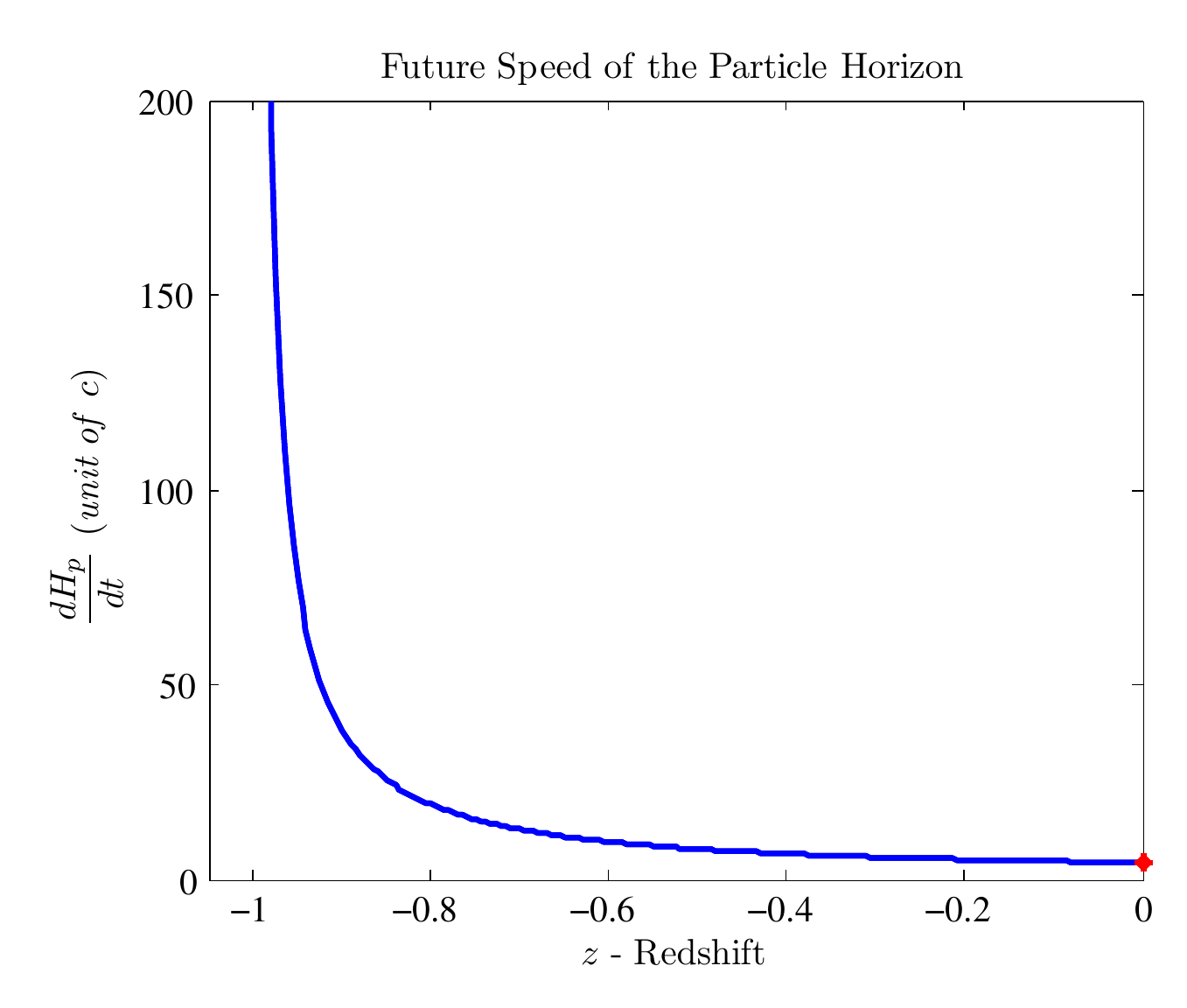} \quad \includegraphics[trim=3.2cm 9cm 4cm 9cm,clip,scale=0.68]{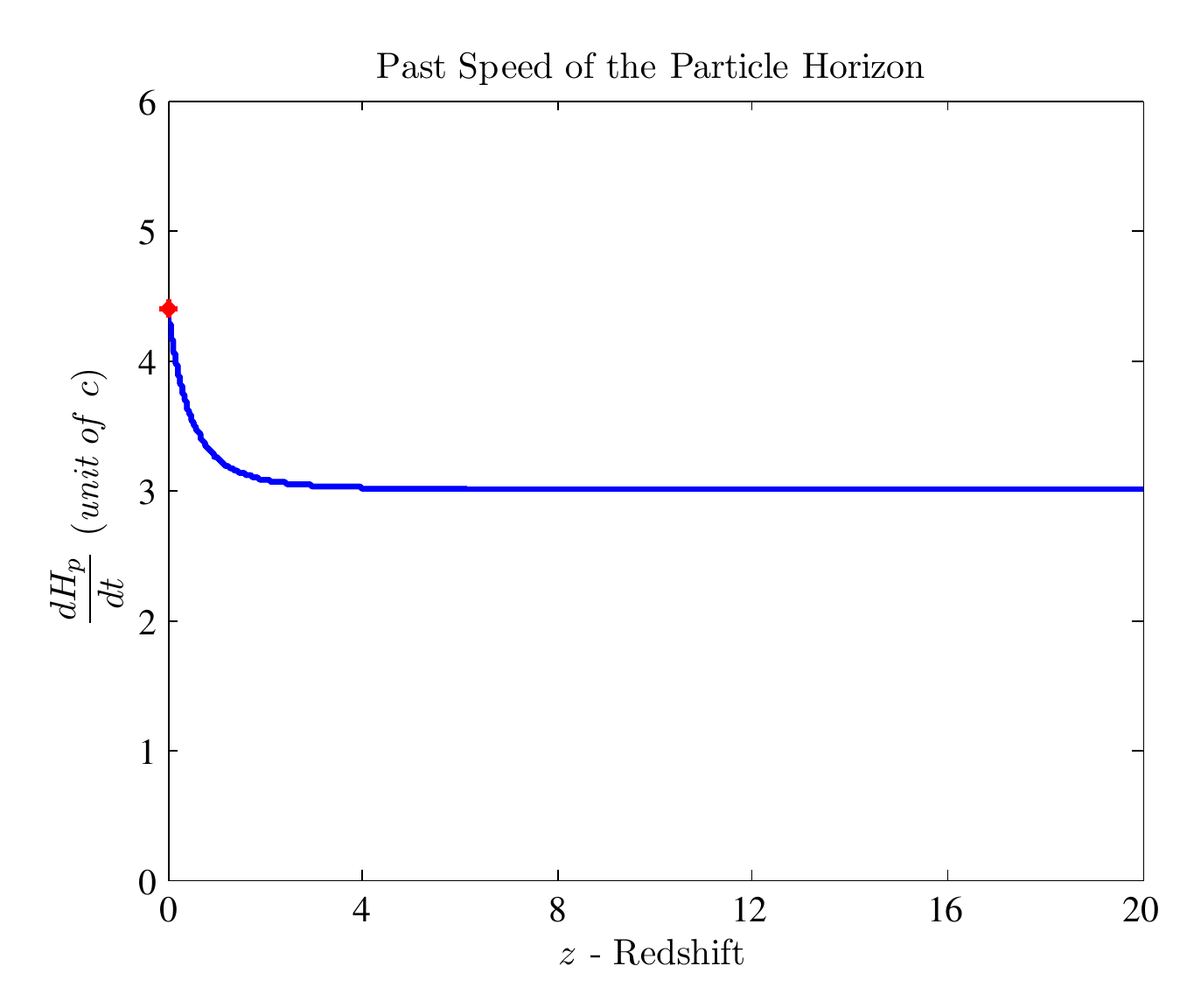}}
      \caption{Velocity of the Particle Horizon. At the origin of the Universe it tends to the constant value \mbox{$3c\simeq2.91\cdot 10^{-14}\mathrm{Mpc/s}$} and increases rapidly to infinite, just as the particle horizon itself.}\label{graphic dHp}
    \end{figure}

    \newpage

  \subsection{Event horizon}
    \begin{figure}[ht!]
      \centering
      \centerline{\includegraphics[trim=3.2cm 9cm 4cm 9cm,clip,scale=0.68]{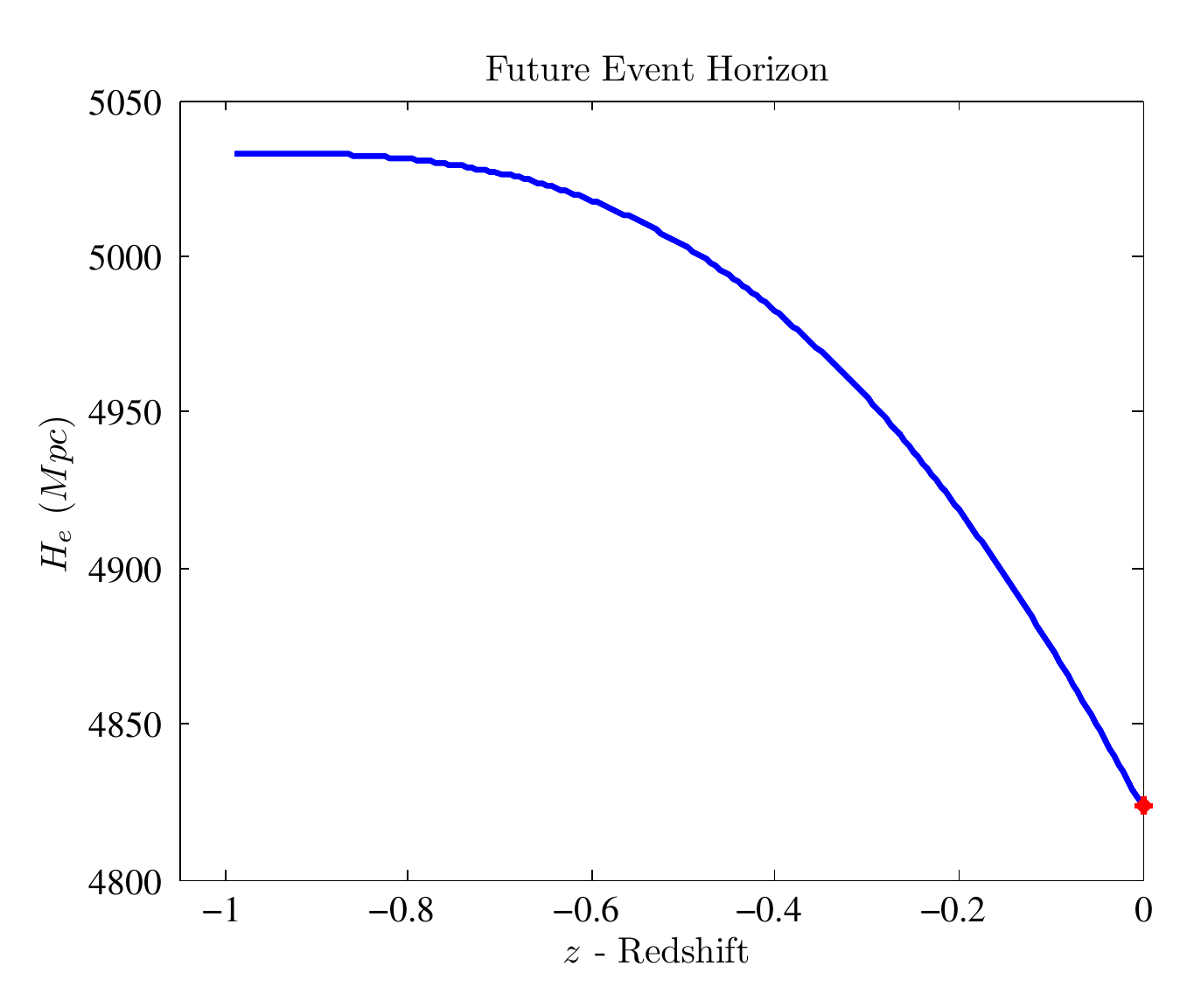} \quad \includegraphics[trim=3.2cm 9cm 4cm 9cm,clip,scale=0.68]{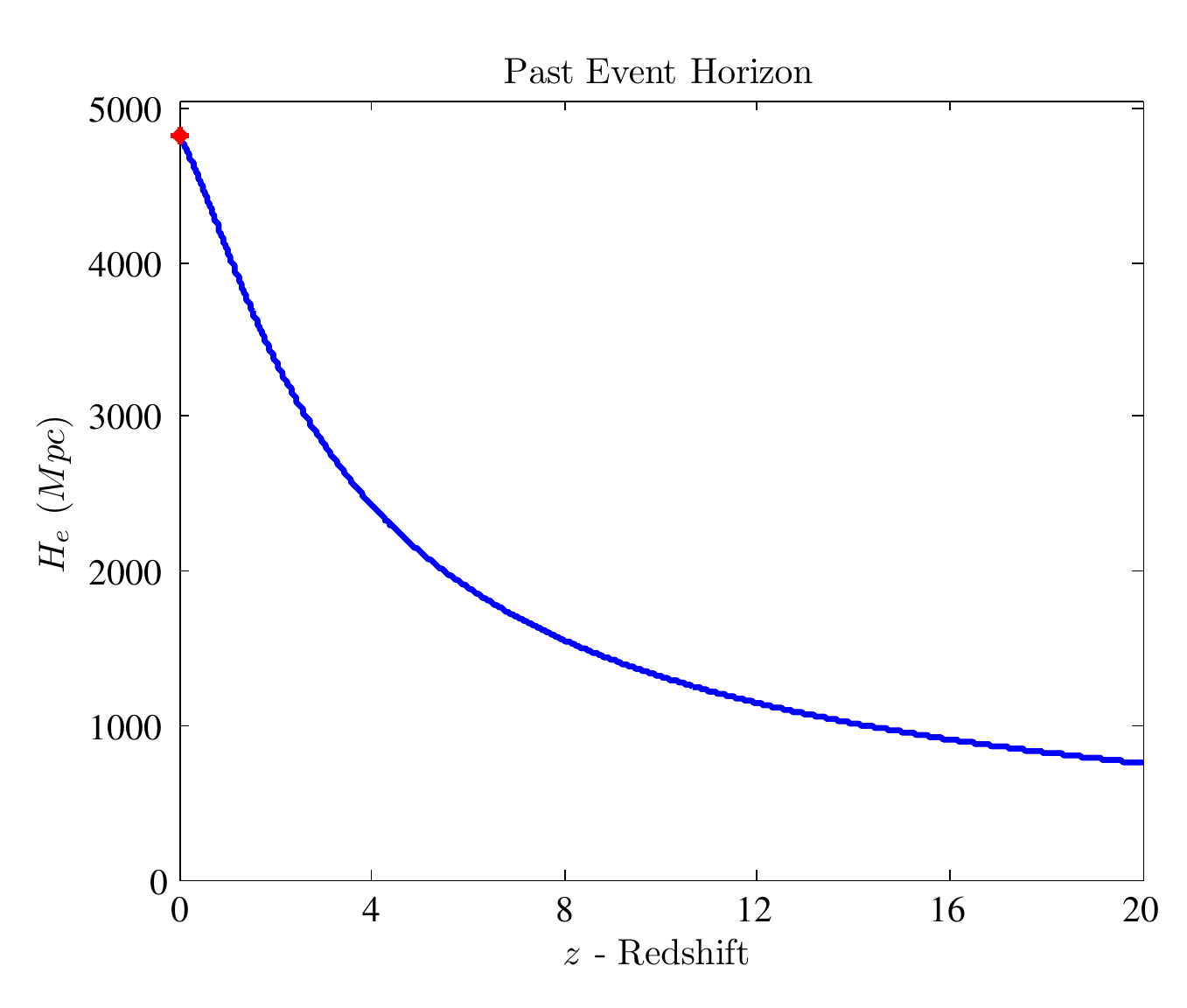}}
      \caption{Event Horizon. At the origin of the Universe, the event horizon tends to zero, then it increases with time, tending at the far future to the constant value \mbox{$c\left(H_0\sqrt{\Omega_{\Lambda0}}\right)^{-1}\simeq5\,033.08\, \mathrm{Mpc}$} with no velocity i.e. with zero slope.}\label{graphic He}
    \end{figure}

    \begin{figure}[ht!]
      \centering
      \centerline{\includegraphics[trim=3.2cm 9cm 4cm 9cm,clip,scale=0.68]{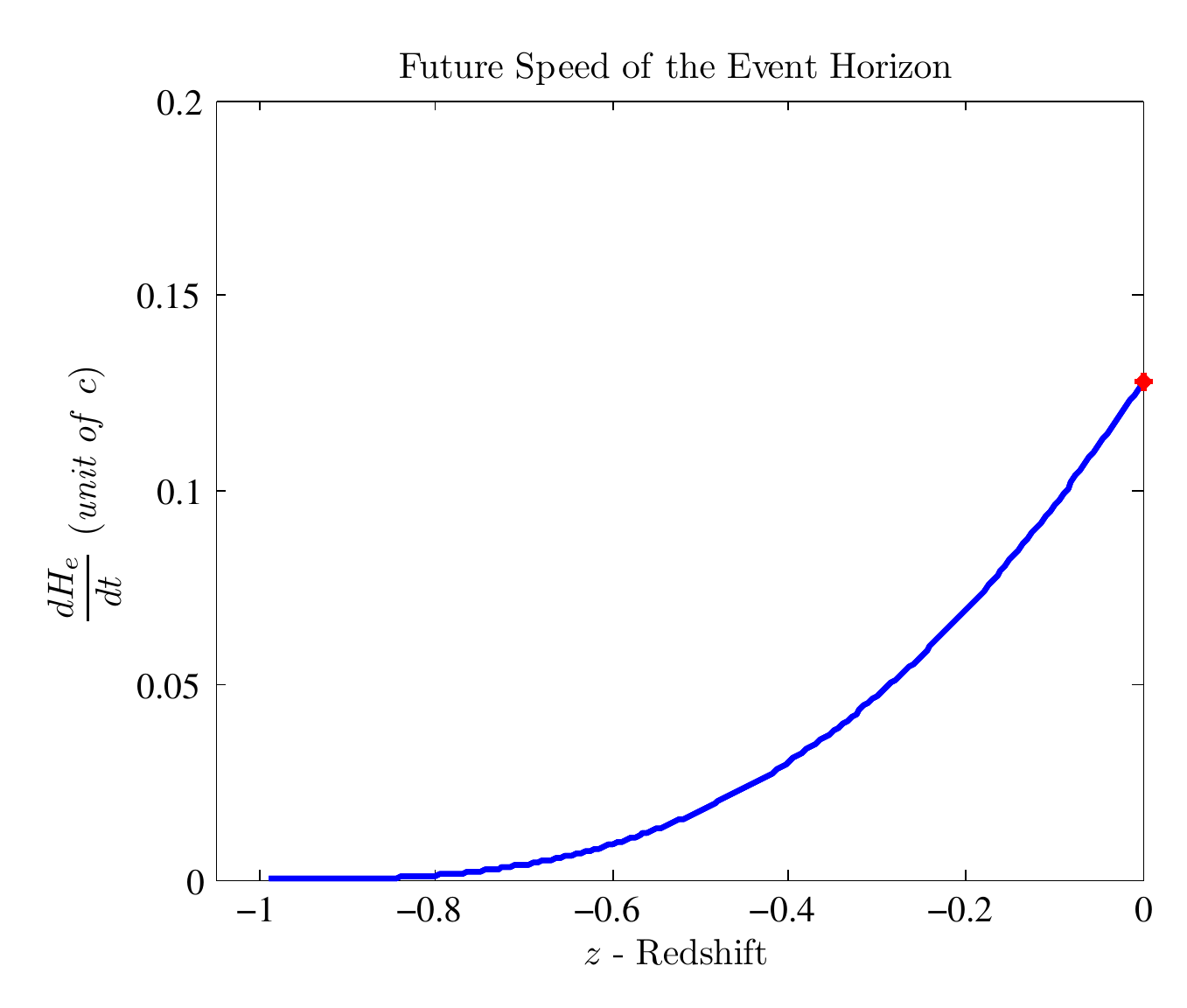} \quad \includegraphics[trim=3.2cm 9cm 4cm 9cm,clip,scale=0.68]{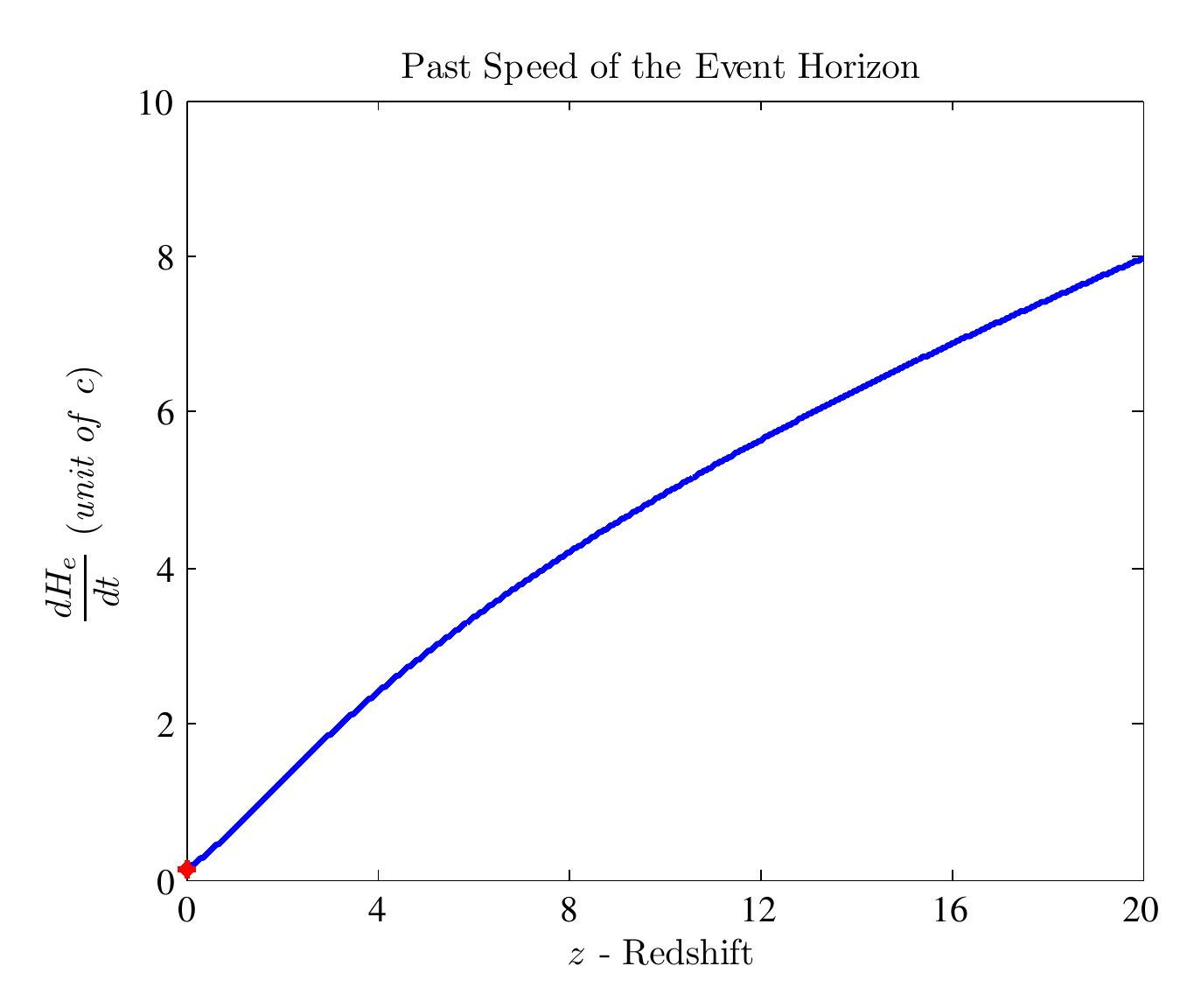}}
      \caption{Velocity of the Event Horizon. At the origin of the Universe it tends to infinite and decrease to zero at the far future with no acceleration (again with zero slope).}\label{graphic dHe}
    \end{figure}
\end{document}